\documentclass[english,twocolumn,aps,pre,10pt]{revtex4-1}
\usepackage[T1]{fontenc}
\usepackage[latin1]{inputenc}
\usepackage{babel}
\usepackage{epsfig}
\usepackage{amsmath}
\usepackage{amssymb}
\usepackage{epstopdf}
\usepackage{overpic}
\usepackage{color}
\usepackage[breaklinks=true]{hyperref}
\usepackage{natbib}
\usepackage{comment}
\hypersetup{colorlinks, citecolor = {blue}, linkcolor= {blue},
 pdftitle = {Extreme Events in a 1d Lattice}}

\begin{document}
\title{Extreme events due to localisation of energy}
\author{Colm Mulhern, Stephan Bialonski, Holger Kantz}
\affiliation{Max Planck Institute for the Physics of Complex Systems, N\"othnitzer Str.~38,
D-01187 Dresden, Germany.}
\email[Corresponding author: ]{mulhern@pks.mpg.de}

\begin{abstract}
We study a one-dimensional chain of harmonically coupled units in an asymmetric anharmonic soft potential. Due to 
nonlinear localisation of energy, this system exhibits extreme events in the sense that individual elements of the chain show very 
large excitations. A detailed statistical analysis of extremes in this system reveals some unexpected properties, e.g., a pronounced 
pattern in the inter event interval statistics. We relate these statistical properties to underlying system dynamics, and notice that 
often when extreme events occur the system dynamics adopts (at least locally) an oscillatory behaviour, resulting in, for example, a 
quick succession of such events. The model therefore 
might serve as a paradigmatic model for the study of the interplay of nonlinearity,
energy transport, and extreme events.
\end{abstract}
\maketitle

\section{Introduction}
\noindent
Extreme events as naturally occurring large deviations of 
a system's state from its normal behaviour have gained considerable interest
during the past years \cite{albeverio, bunde, sornette, ghil}. Many phenomena where one 
likes to speak about extreme events are related to natural disasters,
such as earthquakes, weather extremes, rogue waves. Due to the possible consequences of such events, much work has gone into the 
prediction of where and when they will occur. This allows for better planning in the affected regions; for example, better evacuation 
strategies in coastal areas where tsunamis are a threat. 

However, also in physical models extremes have been discussed, e.g. in microwave
cavities \cite{hohmann}, optical fibres \cite{solli}, superfluids \cite{ganshin}, and semiconductor lasers \cite{bonatto}, to name but 
a few. When studying extremes, four issues are 
in the main focus: the statistics of extreme events, in particular the 
question of how frequently very large events are to be expected; 
temporal correlations between successive events; the underlying mechanism 
for the creation of such large deviations; and the issue of how to 
forecast individual events.

An extreme event in a spatially extended system often presupposes that a significant amount of energy localises in a small area of a 
system (with respect to the 
systems overall size). This problem of energy localisation has received much attention in the area of discrete lattices. The 
genesis of this focus comes from work carried out on what is now referred to as the Fermi-Pasta-Ulam problem \cite{weissert}. One 
of the remarkable discoveries from 
their numerical work on a one-dimensional lattice system consisting of nonlinearly coupled units, was that discrete systems with added 
nonlinearity can localise energy. This led to exciting work 
on \emph{discrete breathers} (also known as intrinsic localised modes) \cite{marin,flach}, which are large amplitude, exponentially 
localised, exact periodic solutions to nonlinear discrete lattice systems. Such solutions are long-lived. In contrast, it is also 
possible for these lattice systems to exhibit transient dynamics which have the characteristic of exponentially localised energy; for 
example, chaotic breathers \cite{cretegny, hennig1}. These solutions typically have finite lifetime with the energy eventually 
dispersing to 
other parts of the system.

The so-called extreme events have been observed in a number systems; for example, in a one- and in a two-dimensional discrete 
nonlinear Schr\"odinger system 
\cite{maluckov1,maluckov2}. The exact mechanism(s) generating extreme events are difficult to pin-point, and thus these studies often 
take the form of a statistical analysis. 
However, one mechanism that is known to account for energy localisation, and indeed a likely candidate as a mechanism that produces 
extreme events, is known as \emph{stochastic localisation}. This mechanism is particularly present in systems where the oscillators 
under investigation are soft, i.e. those whose frequency of oscillation decreases, and the density of states increases, as their 
amplitude increases. Thus, it is favourable (entropically speaking) for such oscillators to seek out regions of phase-space with the 
larger density of states \cite{brown,reigada}.

In this paper we study a model system where we highlight a mechanism 
which might be typical of extremes: the localisation of energy at some few
degrees of freedom. We present a detailed statistical analysis of extremes 
in this system with some 
unexpected properties, e.g., a pronounced pattern in the inter event 
interval statistics. The model being Hamiltonian, the total energy is
conserved. This might be relevant even for real world phenomena, i.e. in dissipative systems, such as 
extreme precipitation events: also there, mass conservation and energy 
conservation play a role, since there cannot be more rain than 
water vapour transported by air, and latent energy has to be released when 
water vapour condensates into rain drops. We also examine the differences and similarities in the dynamics and 
statistical distributions when using two different thresholds that determine the onset of extreme events -- one threshold 
being an amplitude $q=q_{thres}$, and the other being 
a local energy (potential plus kinetic) threshold $E=E_{thres}$. Further, we explore how the system reacts to changes in the coupling 
strength $\kappa$.

\section{System}
\noindent
The system under investigation is a one dimensional conservative system consisting of $N$ coupled units which has a Hamiltonian given 
by

\begin{equation}\label{eq:hamiltonian}
 H({\mathbf{p},\mathbf{q}}) = \sum_{n=1}^{N}\left[\frac{1}{2}p_n^2 + U(q_n) + 
\frac{\kappa}{2}(q_{n+1}-q_n)^2\right],
\end{equation}

\noindent
where $p_n$ and $q_n$ represent canonically conjugated momentum and positions variables, and with local onsite potential

\begin{equation}\label{eq:potential}
 U(q_n) = D(1-\mathrm{e}^{-aq_n})^2
\end{equation}

\noindent
which is known as the Morse potential \cite{Morse}. The parameters $D$ and $a$ control the potential well height and width 
respectively. In the 
limit $q \rightarrow \infty$, $U(q) \rightarrow D$. Taking the following 
dimensionless variables $q'_n = aq_n$, $t' = a\sqrt{D}t$, and $\kappa' = \kappa/a^2D$, the corresponding rescaled equations of motion 
are given by 

\begin{equation}\label{eq:eom}
 \ddot{q}_n = -2(1-\mathrm{e}^{-q_n})\mathrm{e}^{-q_n} + \kappa(q_{n+1}+q_{n-1}-2q_n)
\end{equation}

\noindent
where we have dispensed of the prime notation. This system now has a single parameter 
$\kappa$ which regulates the strength of the coupling between neighbouring units. Throughout we will be using periodic boundary 
conditions and, unless otherwise stated, a chain will consist of 100 units. Moreover, the system's energy density will be equal to 
unity, i.e. $\sum E_n = N$. In general terms, the anharmonic onsite potential can 
be seen to represent intramolecular interactions, while the linear coupling between the units represents intermolecular interactions. 
The particular onsite potential considered here, the Morse potential \cite{Morse}, is regarded as a suitable model for the potential 
energy of a 
diatomic molecule because it can account for the anharmonicity of real chemical bonds, and it implicitly includes the effects of 
dissociation (bond breaking).

To begin, there are two limits that should be 
explored. First, the uncoupled limit $\kappa\rightarrow 0$ will exclude the possibility 
of energy propagating through the chain, with each chain unit maintaining its initial energy. Thus, if some chain unit possesses 
energy $E_n \geq 1$ then this unit can travel to infinity in the positive $q$ direction. The other limit, $\kappa \rightarrow \infty$, 
results in infinitely strong interactions between neighbouring units. For systems with finite energy, the only possible chain 
configuration is a uniform rod-like state where the dynamics of each unit is the same. Any deviation from this rod-like state would put 
an infinite amount of energy into the harmonic springs, something that is clearly not possible in the finite energy case. Importantly, 
the limits $\kappa=0$ and $\kappa=\infty$ are integrable. The interesting collective dynamics, and consequently the interesting 
statistics, come in the nonintegrable case, i.e., for finite nonzero values of the coupling strength.

In this paper we are primarily concerned with extreme events. An extreme event is said to have occurred at some chain location (unit) 
when this unit has an 
amplitude greater than a predefined threshold, that is when $q_n\geq q_{thres}=4$. Here, we will also look at an energetic threshold 
defined 
in terms of local energy. An extreme event in terms of local energy occurs when the sum of the potential plus the kinetic energy of a 
chain unit crosses a threshold, that is $E_n\geq E_{thres}=3$. The coupling part of the potential energy is shared evenly between 
neighbouring units.

We expect 
that small changes in 
the above thresholds will not have qualitative implications for the statistical results that will be discussed later. Of course, other 
thresholds will naturally produce different results, and there is no perfect way to define the thresholds. Indeed, we are fully aware 
that the statistics discussed later depend on how the thresholds are defined. However, the thresholds chosen 
here seem reasonable. The $q=4$ threshold, being well beyond the inflection point of the (Morse) potential at $q \approx 0.693$, 
corresponds to this potential having 
a gradient close to zero, with energy (excluding the coupling and kinetic energies) $E\approx1 = U(\infty)$. The $E=3$ 
threshold means that a unit will have energy that is at least 3 times the energy density of the system. Moreover, choosing a fixed 
threshold 
allows for a direct comparison between the statistics for different coupling strengths. Thus, an extreme event occurs when a unit 
either crosses a particular point in space, or when it attains a sufficient amount of energy. These thresholds are quite different, 
and importantly, a threshold crossing in space or energy does not necessarily imply a crossing in the other; that is, a high 
concentration of energy at a particular chain unit, does not imply a large amplitude excitation, and vice versa.

\section{Dynamics}
\begin{figure}
\includegraphics[trim=0cm 0cm 0cm 2.6cm,clip=true,width=0.45\textwidth,height=9cm]{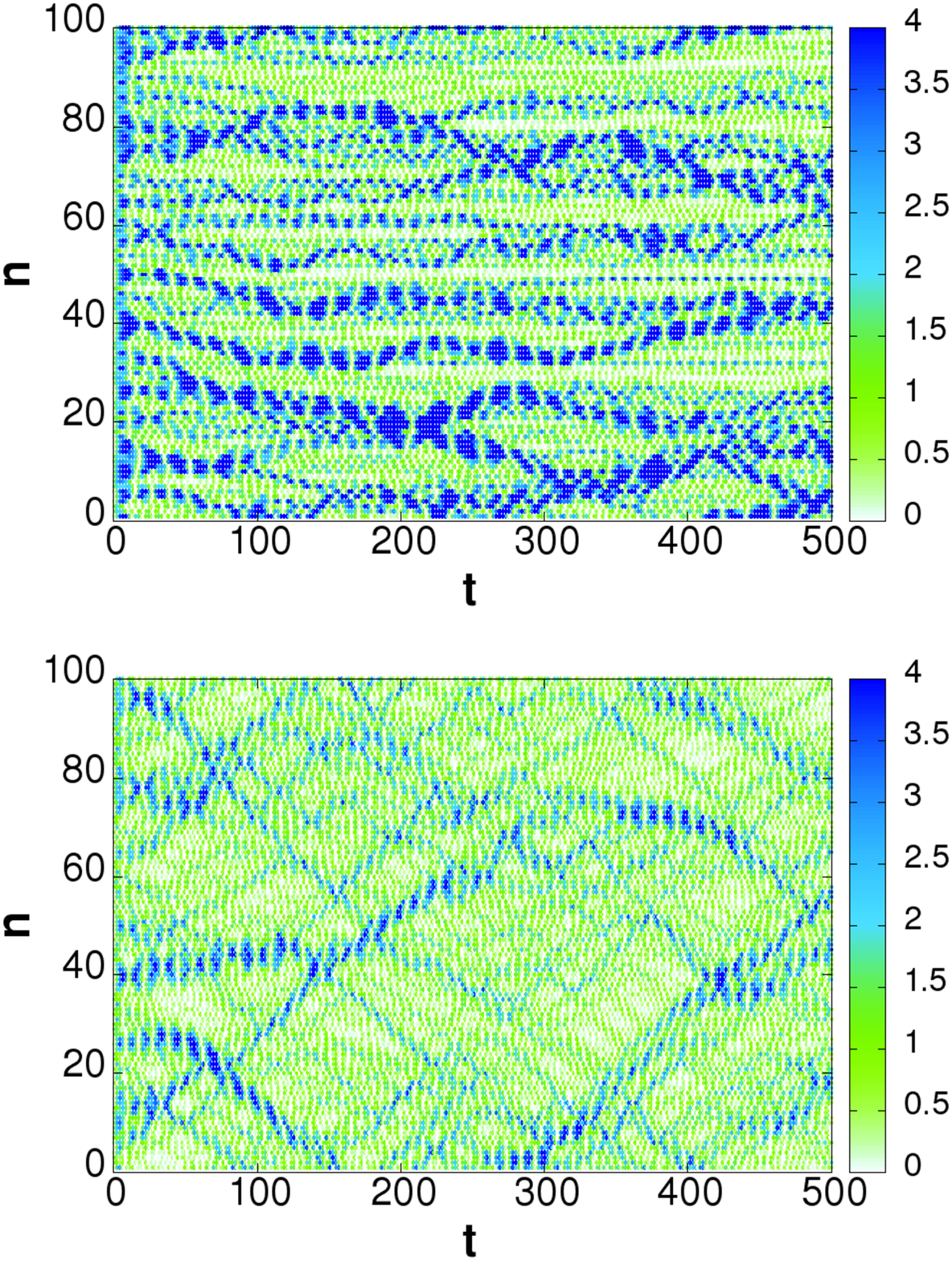}
\caption{Color online. Heat maps showing the evolution of the amplitudes $q_n(t)$ for 100 coupled units, as a function of time, for two 
different 
values of the coupling 
strength: $\kappa=0.1$ (top), $\kappa=0.4$ (bottom). All amplitudes $q_n(t)\geq 4$ are coloured blue. Here $t$ denotes time, and $n$ 
the 
location of the units on the chain. The oscillating island structures are clearly visible.}\label{fig:heat}
\end{figure}

With a view to the later discussion on \emph{extreme events} it is worthwhile to look at the dynamics of the system under 
investigation. For this, we inject kinetic and potential energy randomly into the system such that $E_n(0)=1$ for all $n$, and 
subsequently study the 
evolution of this initial excitation. Fig.~\ref{fig:heat} contains heat maps which allow one to observe the evolution of the 
amplitudes $q_n$ through time. Interestingly, it can be seen that chain units can undergo large fluctuations in their 
amplitudes. In the corresponding energy heat maps (not shown) it can be seen that the energy can become concentrated (localised) at 
certain chain 
locations. The localised island-like structures, in amplitude and energy, that emerge are able to move along the chain; the speed 
of this mobility seems to depend on the 
strength of 
the coupling between neighbouring units, the number of units involved in the structure, and on the energy contained in the structure. 
This mobility allows for the interaction of
two or more localised structures resulting in various consequences. A number of possible outcomes from these interactions include, the 
creation of a 
larger structure through the merging of smaller ones, a dispersion of the energy contained in the structures, two structures reflect 
off one another, etc \cite{cuevas1}. This presents an algorithmic challenge when trying to monitor the time 
evolution of these localised structures, as it is unclear what happens to the individual structures when they collide with others. 
Looking more closely at the individual localised structures, it can be seen that they are oscillating in time. This can be observed in 
Fig.~\ref{fig:heat} by noticing that the localised structures with high amplitudes are intermittently interrupted by periods 
of low 
amplitude, i.e. in the heat maps the colours of the localised structures alternate between green and blue. 

\begin{figure}
  \includegraphics[scale=0.3]{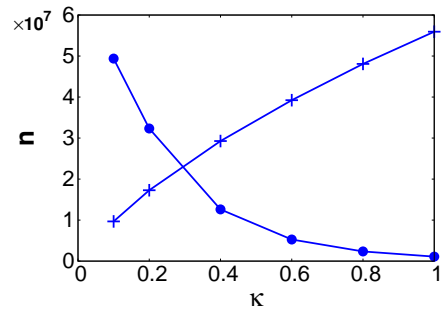}
\caption{Number of $q_{thres}$ (filled circles) and $E_{thres}$ (crosses) threshold crossings as a function of the coupling 
strength $\kappa$.}\label{fig:number}
\end{figure}

The system dynamics also react in interesting ways when the coupling strength is changed. With increasing coupling strength energy can 
disperse more freely along the chain. Thus, one can see localised structures moving over a significant portion of the chain in short 
periods of time. On the other hand, increasing the coupling impedes the formation of large amplitude localised structures. This is 
because more energy is required in the harmonic springs. This, in turn, means that there will be fewer $q_{thres}$ threshold crossings 
with 
increasing coupling strength. Fig.~\ref{fig:number} shows the number of $q_{thres}$ threshold crossings for the chain units 
as a function of the coupling strength $\kappa$ for an ensemble of 1000 initial conditions, and simulation time per initial condition 
$t=5\times10^4$ ($\approx 1.1\times 10^4\times T_0$ with 
$T_0 = \sqrt{2}\pi$ being
the period duration of harmonic oscillations about the potential
minimum). The curve representing the 
number of threshold crossings shows a tendency, as $\kappa$ is increased, to approach a constant (see below). However, one 
needs to consider that in the uncoupled case $\kappa=0$, for the initial 
conditions considered here, each unit in the chain will cross the threshold once before escaping to infinity. This results in 
$100\times 1000 = 10^5$ (number of units times number of initial conditions) threshold crossings. Therefore, moving into the coupled 
regime, one must conclude that there is a transition which sees a dramatic increase in the number of extreme events in $q$ before this 
behaviour reverses and there is a decay in the 
number of events. One may be inclined to think that there is a singularity at $\kappa = 0$ causing a large jump in the number of events 
as one moves away from this limit. However, the finite simulation time, together with the very weak coupling that allows chain units to 
travel large distances almost unhindered, more likely points to a gradual smooth increase in the number of events as $\kappa$ is 
increased from $\kappa=0$.

Dynamical behaviour, analogous to that discussed above, can be observed for energy heat maps (not shown), where the local kinetic plus 
potential 
energy for each chain unit are monitored. However, the oscillating nature of the localised structures is less apparent. This is due to 
the fact that energy in a unit can stay the same even when its amplitude drops, as the potential energy can be converted into kinetic 
energy. The consequences for the number of $E_{thres}$ threshold crossings  as $\kappa$ is increased can be seen in 
Fig.~\ref{fig:number}. Notably, the number of crossings increases 
with the 
coupling strength. Again, it is easy to deduce the number of crossings in the $\kappa = 0$ case: without coupling to other units, 
no unit is able to attain the additional required energy to cross the threshold, and so there will be no energetic extreme 
events. Further, while large amplitude 
excursions by the chain units become less likely with increased $\kappa$, the same cannot necessarily be said for the local energy. 
That is, localisation of energy is likely to be enhanced with increased coupling strength. Of course this is true only up to some 
critical $\kappa$ value, as the chain will approach the rod-like configuration (mentioned earlier) due to energetic constraints for 
high 
values of $\kappa$. In fact, this critical value is almost certainly $\kappa=\infty$ (discussed below), as for any finite coupling 
strength, there is nothing that forbids small amplitude deviations between neighbouring units. This presents a clear difference between 
the two chosen thresholds. That this same process (of increasing $\kappa$) 
allows for more frequent energetic extreme events, in contrast to the $q_{thres}$ extreme events, is interesting. It would 
seem that energy still localises efficiently without the need for the chain units to attain large amplitudes.

Finally, it is 
worth discussing the number of $q_{thres}$ and $E_{thres}$ threshold crossings in the case $\kappa=\infty$. We reiterate, that in this 
case, the only possible chain configuration is that where the chain is rod-like and each unit undergoes exactly the same dynamics. This 
means that zero energy is contained in the coupling part of the potential. Now we have a chain acting (effectively) like a single 
unit, with energy $E=1$. This is exactly the $\kappa=0$ case, and so the number of threshold 
crossings, $q_{thres}$ and $E_{thres}$, are 
exactly the same in both the $\kappa=0$ and $\kappa=\infty$ cases: $n(q_{thres}^{\kappa=0})=n(q_{thres}^{\kappa=\infty})=10^5$, and 
$n(E_{thres}^{\kappa=0})=n(E_{thres}^{\kappa=\infty})=0$. Turning our attention to the curves contained in 
Fig.~\ref{fig:number}, the $q$-curve will likely continue with an exponentially slow decay, with increasing $\kappa$, towards its 
$\kappa=0$ value. On the other hand, the $E$-curve must at some point reverse its behaviour and begin a descent towards zero with 
increasing $\kappa$.

\section{Energy Localisation}
To give a quantitative characterisation of how well the energy localises for various coupling strengths, we introduce the following 
quantity \cite{cretegny}:

\begin{equation}
 L_N(t) = \Bigg \langle N\dfrac{\sum_{i=1}^{N}E_i^2(t)}{\big(\sum_{i=1}^{N}E_i(t)\big)^2}\Bigg \rangle
\end{equation}

\noindent
where $N$ is the number of units in the chain, and $E_i$ is the local potential plus kinetic energy contained at chain location 
$i$. The angled brackets indicate an ensemble average over $100$ initial conditions. For full equipartition of energy, that is 
$E_i=E/N$ for all $i$ and total system energy $E$, $L_N=1$. Conversely, if all the system energy is concentrated at some single chain 
location $n$, that is $E_n=E$ and $E_i=0$ for all $i\neq n$, then $L_N=N$.

\begin{figure}
\includegraphics[height=5.3cm,width=7.2cm]{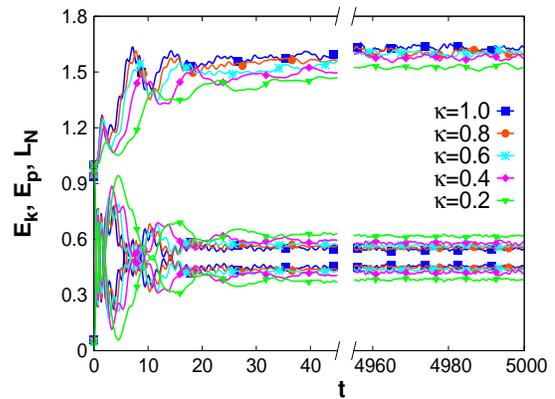}
\caption{Color online. The curves at the top of the figure show the evolution of the localisation measure $L_N$, the middle curves are 
the 
potential energy, and the bottom curves are the kinetic energy. For each coupling strength (see key) the curves have been obtained 
using the same ensemble of 100 trajectories. Note that there is a break in the axis at $t=45$.}\label{fig:locM}
\end{figure}

The results for various values of the coupling strength $\kappa$ are contained in Fig.~\ref{fig:locM}. Starting from a state of full 
equipartition of energy a clear trend emerges, regardless of the coupling strength. The early evolution of $L_N$ is characterised by an
overall increase from its initial value of $E_N=1$. Subsequently, $L_N$ settles into its long term behaviour plateauing close to a 
maximal value. It should be noted that the evolution of $L_N$ for individual trajectories is much more erratic with numerous peaks 
which go significantly above the average values. Also worth remembering is that $L_N$ remains constant in the case of uncoupled units 
-- that means $L_N=1$ for the initial set up considered here. While the measure of localisation might appear low, $L_N\ll N$, even 
those events which show large amplitude excitations need to be considered with respect to a high energy background. This will be 
discussed in the coming paragraphs.

The evolution of $L_N$ can be related to the corresponding evolution of the average local kinetic and potential energies (see also 
Fig.~\ref{fig:locM}). In particular, a transient period, where the average kinetic and potential energies oscillate erratically, sees 
an 
increase in $L_N$, i.e. an increased localisation of energy. Afterwards, the soft nature of the underlying Morse potential emerges 
with the average potential energy becoming higher than the corresponding kinetic energy. Moreover, these energies now undergo small 
oscillations about some constant value which is dependent on the coupling strength. At the same time, $L_N$ has settled into his long 
term behaviour.

\begin{figure}
\begin{overpic}[height=5.0cm,width=7.0cm]{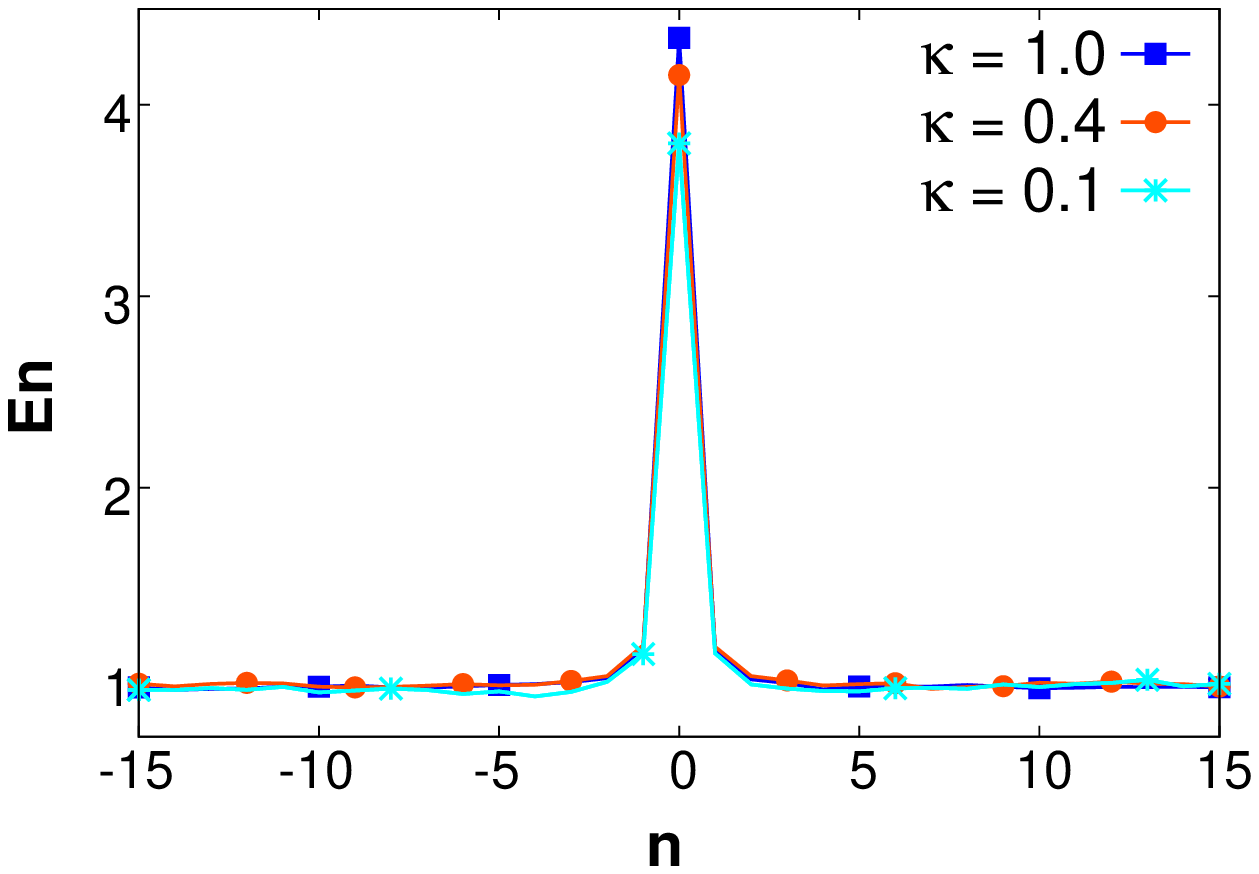}
\put(11,40){\includegraphics[height=2.0cm,width=2.83cm]{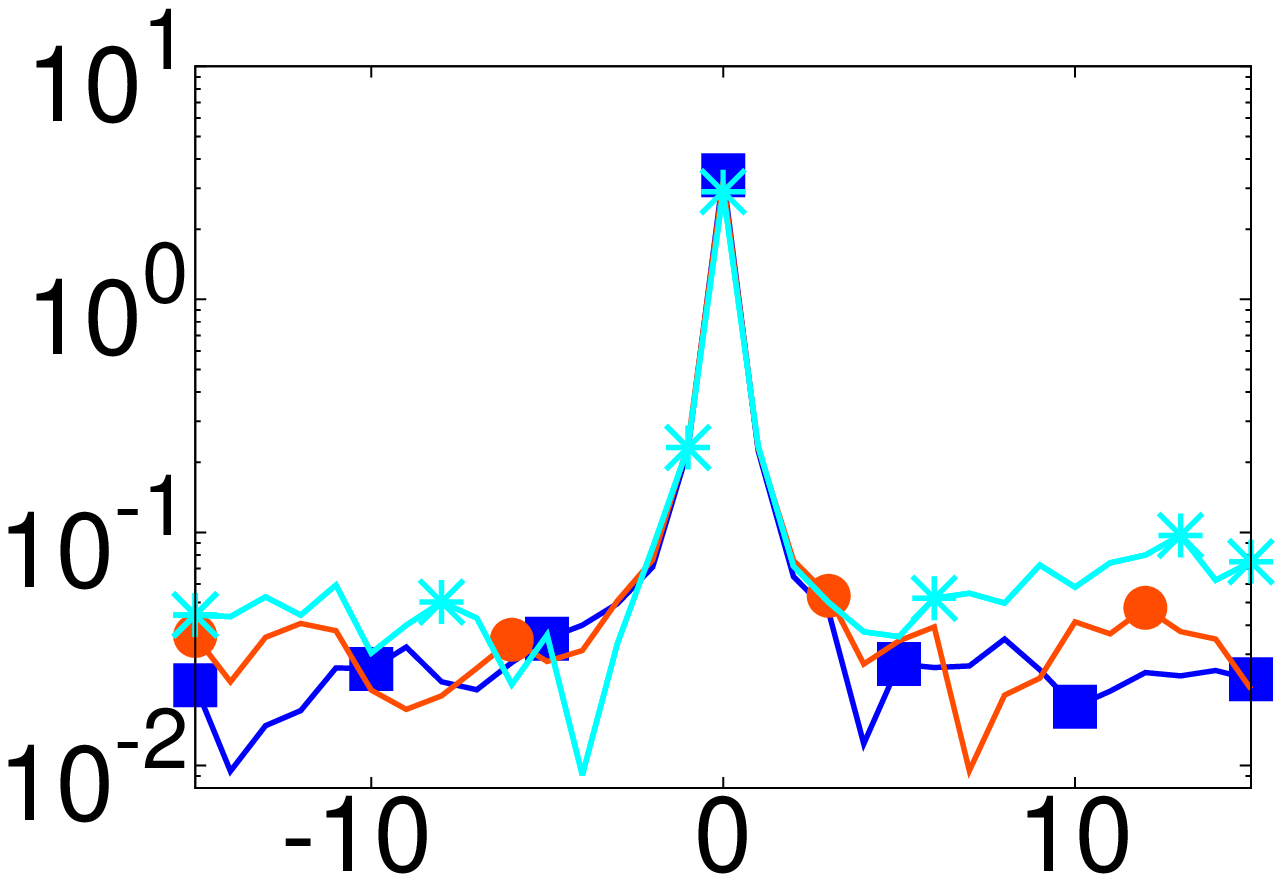}}
\end{overpic}
\caption{Color online. The average energetic excitation pattern in the vicinity of the location of 
the maximum $i_m$, when the excitation exceeds the $E_{thres}=3$ threshold. 
The logarithmic representation (inset) shows that it is, to a good approximation,
the superposition of some constant background and a symmetric exponential
part, hence we observe exponential localisation in space.}\label{fig:profiles_q}
\end{figure}

To see how the energy localises for extreme events we show in Fig.~\ref{fig:profiles_q} an 
excitation pattern (with the positions centred at $n=0$) obtained by averaging over the maximum amplitude of all energetic extreme 
events, and 
its 30 nearest neighbours, for a collection of $2\cdot 10^5$ extreme events. To be 
precise, let $i_m(t')$ be the position of the largest 
excitation at the time $t'$ of the extreme event, with $E_{i_m}(t')>E_i(t')$ $\forall i \neq i_m$: we find the mean of $2\cdot 10^5$ 
such excitations with 
the 
position $i_m(t')$ shifted to 
zero. Whereas the linear 
scale shows a strong concentration of the 
excitation pattern at 0, which is the position of 
the maximum, a logarithmic representation, after subtracting the background $E_{min}$ from each curve, shows evidence for exponential 
localisation. Qualitatively similar results have been obtained for the averaged $q$-amplitude excitation pattern (not shown).

Returning to the localisation measure $L_N$, we find that the curves shown in Fig.~\ref{fig:profiles_q} fall in the range $L_N \in 
(1.2,1.35)$. However, when the constant background $E_{min}$ is removed from each curve, $L_N$ then falls in the range $L_N \in 
(12.8,19.2)$ - that is $1 \ll L_N < N = 31$. This demonstrates that all energy localisation must be considered in terms of a high 
energy background, and thus explains why the curves for $L_N$ in Fig.~\ref{fig:locM} appear to suggest that energy does not 
localise efficiently in this system.

\section{Statistics of Threshold Crossings}\label{sec:stats}
To gain further understanding of threshold crossings, going beyond an analysis of the internal dynamics which produce such crossings, 
it is useful to examine their statistical properties. We consider an event to be  \emph{extreme} when a chain unit crosses 
over the threshold $q_{thres}=4$ ($E_{thres}=3$). This event begins with a threshold crossing, i.e. $q_n(t_1)>q_{thres}$, and ends when 
this same units recrosses the threshold, i.e. $q_n(t_2)<q_{thres}$, where $t_1$ and  $t_2$ denote the times of the threshold crossings 
(and similarly for $E_n$). Fig.~\ref{fig:stats_q} contains a number of distributions detailing various 
statistical aspects of threshold crossings, for $q_n$ and $E_n$ respectively, and in particular how these distributions change as a 
function of the coupling strength $\kappa$. 

For the simulations we used 1000 initial conditions, each of which was chosen randomly, but with the restriction that 
$E_n(0)=1$ for all $n$. The simulation time for each initial condition was $t=5\times10^4$ ($\approx 1.1\times 10^4\times T_0$ with 
$T_0 = \sqrt{2}\pi$). The results will be discussed below.

\subsection{Extreme Events: \texorpdfstring{$q$}{q}}
\begin{figure}
\includegraphics[scale=0.32]{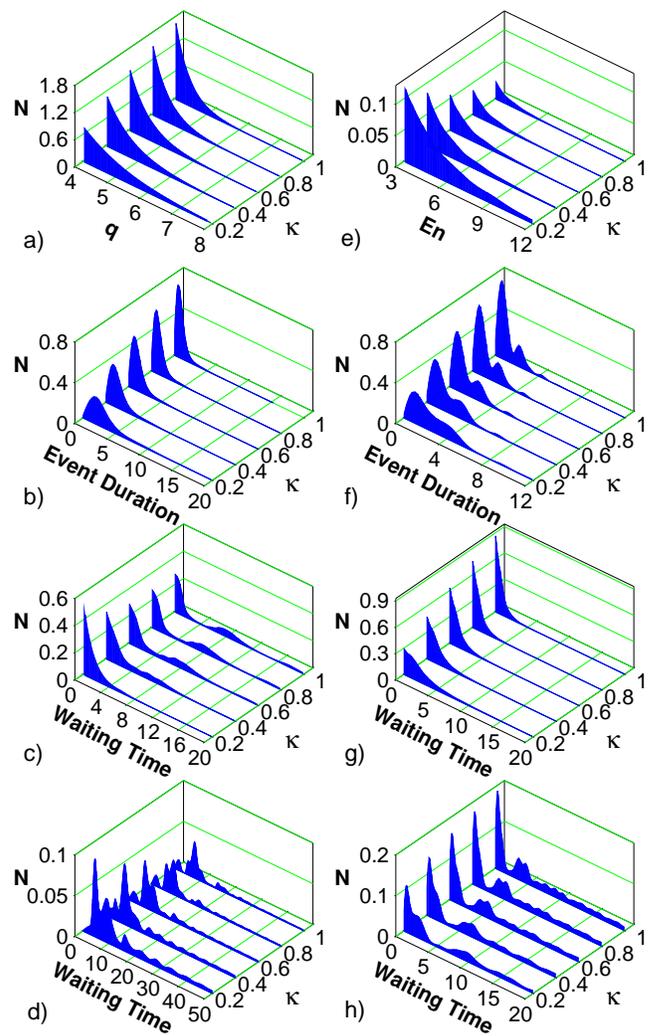}
\caption{Threshold crossing statistical distributions for $q$ (left) and $En$ (right). Row 1) maximum amplitudes for each event 
considered extreme; 2) event 
duration times for events considered extreme; 3) waiting times between events happening anywhere on the chain; 4) waiting times between 
consecutive events happening at the same chain location. More details in text.}\label{fig:stats_q}
\end{figure}

\begin{figure}
\includegraphics[scale=0.22]{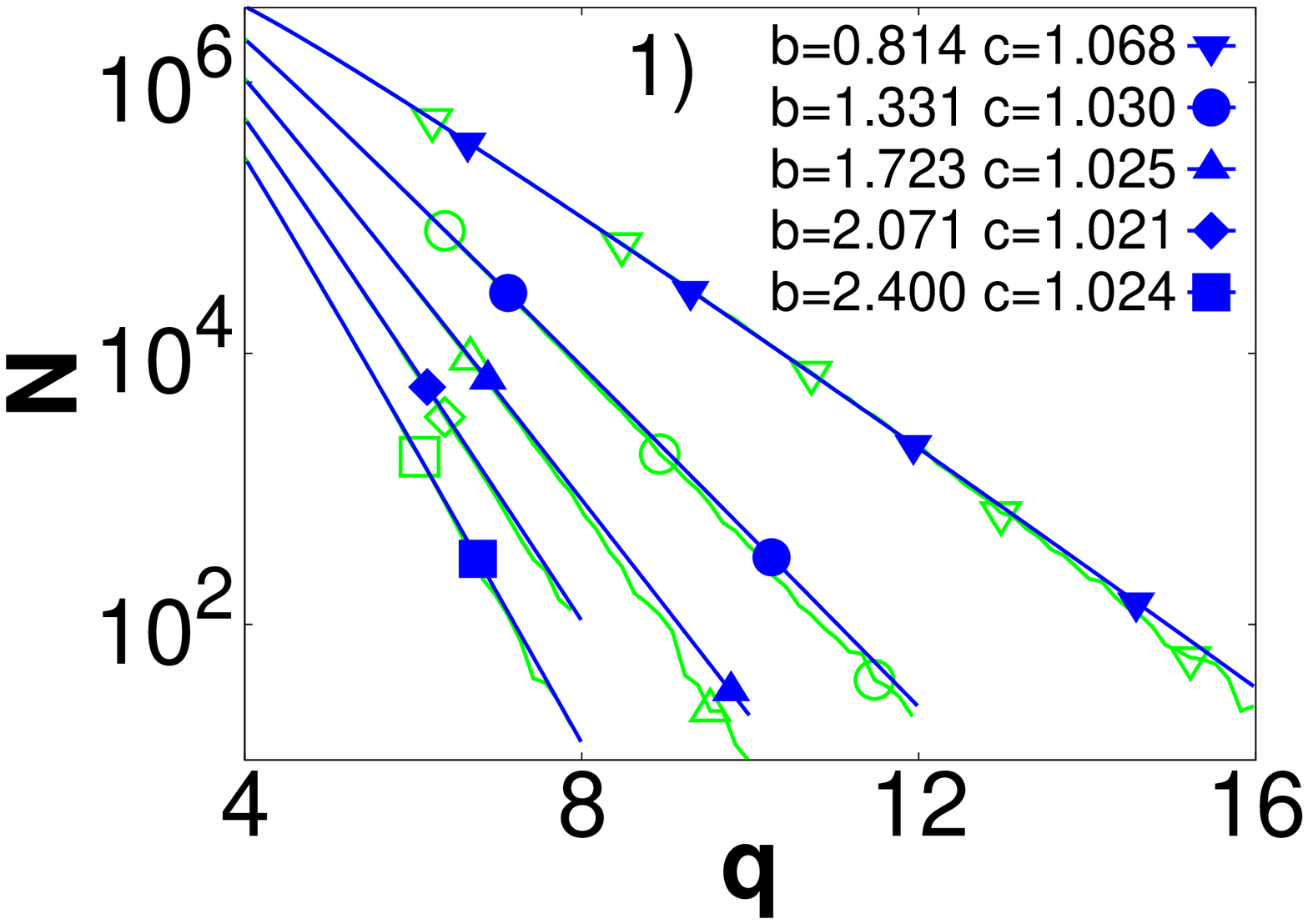}
\includegraphics[scale=0.22]{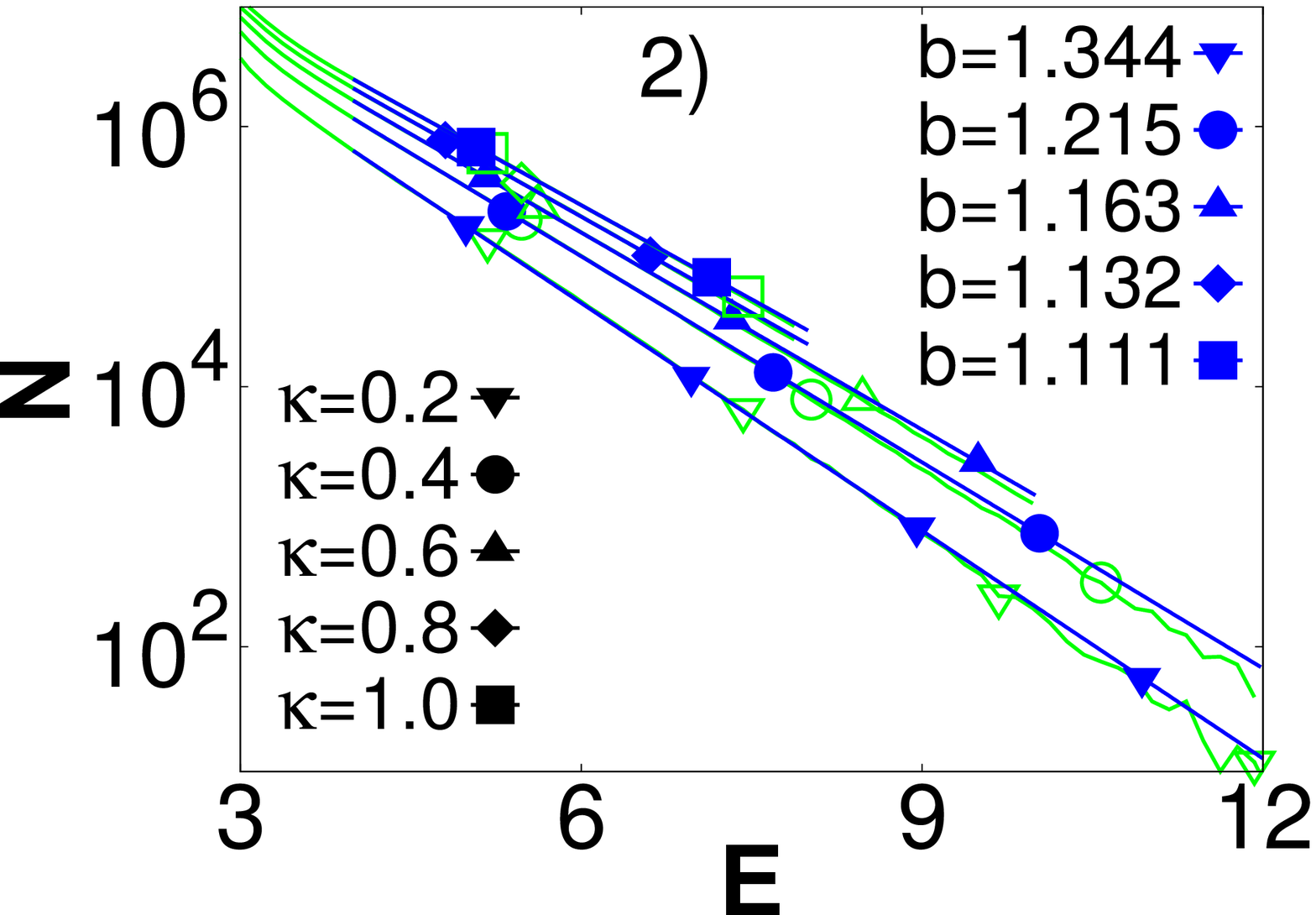}\\
\includegraphics[scale=0.22]{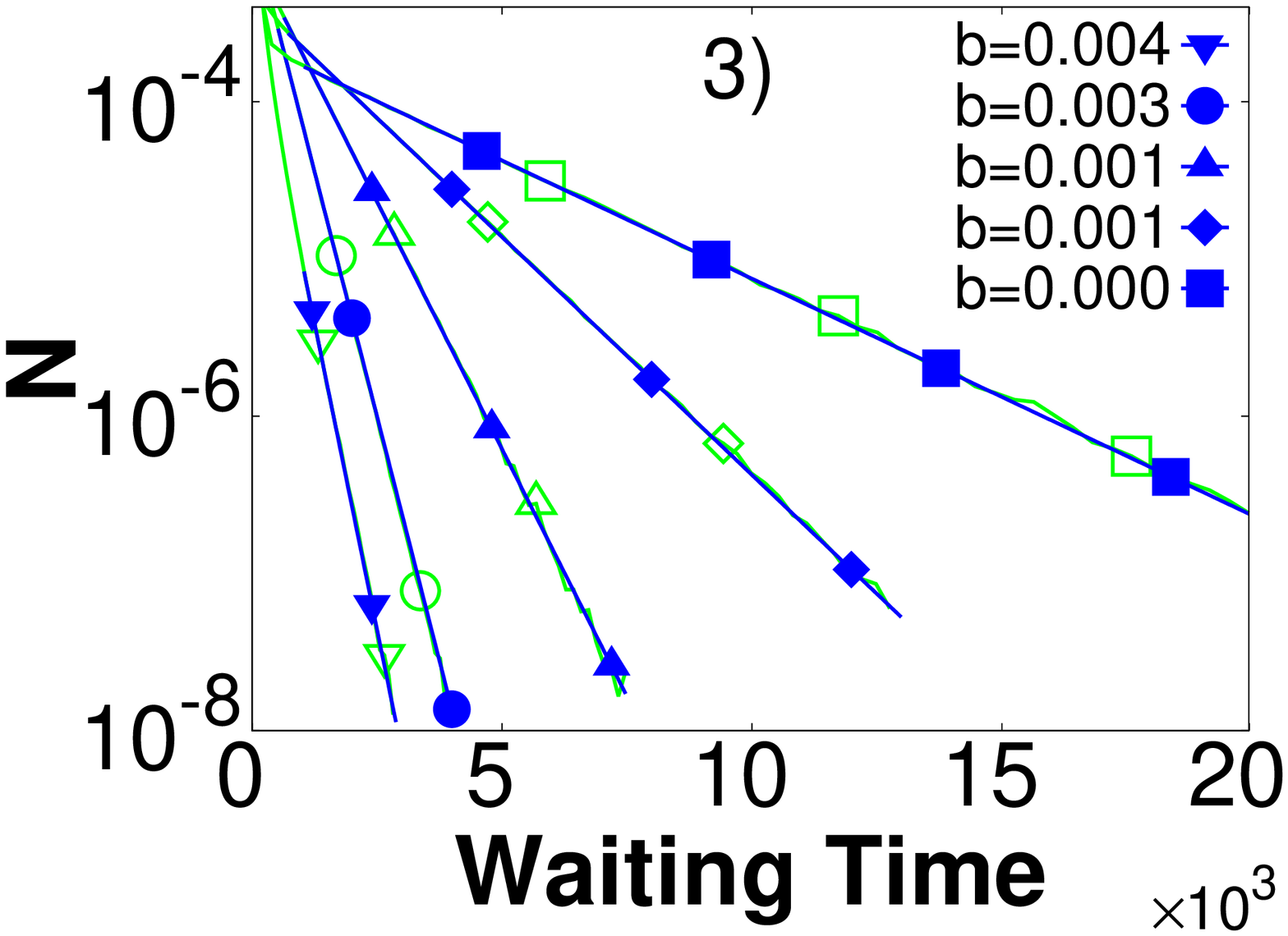}
\includegraphics[scale=0.22]{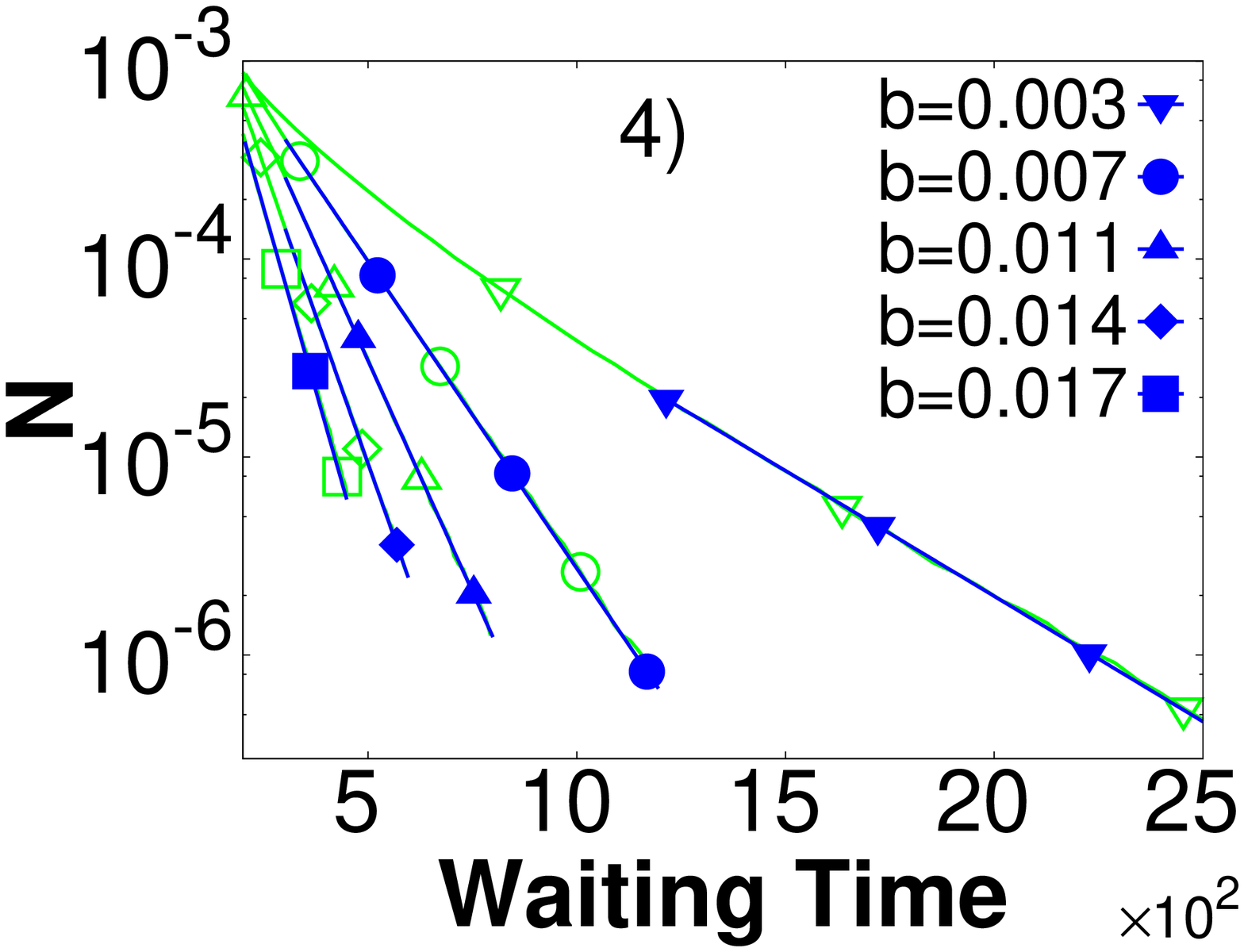}\\
\caption{Color online. Lin-log plots for some of the results contained in Fig.~\ref{fig:stats_q} -- panels 1),2),3),4) correspond 
respectively to 
panels 
a),e),d),h) from Fig.~\ref{fig:stats_q} (empty symbols - further details are contained in the key). Also shown are fitted 
curves 
obtained using the function $f(q)=a\exp(-bq^c)$ (filled symbols). The corresponding values $b$ and $c$ are contained in the key ($c=1$ 
in 
the cases where $c$ is not explicitly given in the key), while $a$ is a constant representing the bin height at 
$q=4,E=3$. The coupling strength related to each symbol is contained in the key in the top right panel.}\label{fig:log3}
\end{figure}

Looking first of all at the distribution for the maximum amplitudes of each extreme event (Fig.~\ref{fig:stats_q}a), a clear trend 
emerges. 
Each 
distribution in $\kappa$ appears to 
exhibit, at least to a very good approximation, an exponential decay. That is, the probability of 
an extreme event attaining ever higher amplitudes decreases exponentially, independent of the value of the coupling strength. 
In fact, after fitting these curves to determine the exact nature of the decay, it was established that the exponential function 
$f(q)=a\exp(-bq^c)$ is an appropriate fit, where the parameters $a$, $b$, and $c$ are constants. The results for this fit are 
contained in Fig.~\ref{fig:log3}a. Evidently, the coupling strength $\kappa$ influences the value of the corresponding exponents: an 
increase in $\kappa$ sees also an increase in the exponent $b$. This manifests itself in the narrowing of the distributions 
seen in Fig.~\ref{fig:stats_q}a. A simple explanation for this increase in $b$ is that stronger couplings inhibit 
a unit's ability to attain large amplitudes. In each case the parameter $c$ remains close to $c=1$, and even shows some convergence 
to this value. However, fitting the curves with $c=1$ (not shown) produced significant errors, compared to the the case $c\neq 1$, 
especially in the tails of the corresponding distributions.

This fitting now enables some predictive power. It is possible now to predict the probability of an extreme event, with a certain 
amplitude (or greater than a given amplitude), taking place. In addition, one can compute the expected value of an extreme event 
$E(q_{>thres}) = \int_4^{\infty}dq[q\times a\exp(-bq^c)]$. For $\kappa = 0.2,0.4,1.0$ one obtains $E(q_{>thres}) = 5.06,4.69,4.33$ 
respectively.

The distributions for the duration of the extreme events (Fig.~\ref{fig:stats_q}b) shows that events typically last no longer 
than $t=5$ times units. The shape of this distribution is maintained, regardless of the considered coupling strengths. 
However, the peak tends towards zero with increasing coupling strength. Moreover, the height of the peak increases, meaning that short 
duration events become more likely than the long duration events. From this result, it is not apparent that some of the localised 
structures have oscillatory behaviour, which 
can result in multiple extreme events in quick succession. One could argue that this succession of events instead accounts for a single 
extreme event, thus increasing the overall duration of an event. However, this is extremely difficult to resolve in such a statistical 
analysis.

Since there is no unique way of defining the waiting time between extreme events we consider two different definitions which, 
interestingly, produce qualitatively different 
results. The first considers extreme events when the consecutive  events can happen anywhere on 
the chain. More precisely, if $q_n(t_1)>q_{thres}$ and $q_m(t_2)>q_{thres}$ where $t_1<t_2$ 
denote the start times of consecutive events, at chain locations $n$,$m$ respectively, then the waiting time between events is 
simply $t_{wait} = t_2 - t_1$.

Secondly, we look at the waiting times between the end of one extreme event and the beginning of the next extreme event 
that occur at the same chain location. To be more precise, if $q_n(t_1)<q_{thres}$ and $q_n(t_2)>q_{thres}$, where $n$ denotes the 
chain 
location, $t_1$ marks the end of one event, and $t_2$ marks the beginning of the next, then the waiting time between these events 
is defined as above. Analogous definitions in terms of the energetic threshold can also be made.

The distributions obtained when considering the first definition appear in Fig.~\ref{fig:stats_q}c. They show a tendency for 
the waiting times to increase with increasing $\kappa$. This is represented in the results by a lowering of the large peak and a 
broadening of the distribution. 
Interestingly though, additional peaks, with amplitudes increasing with $\kappa$, emerge. To explain this, we should again refer to 
the heat maps contained in Fig.~\ref{fig:heat}. For low $\kappa$ values there can simultaneously coexist multiple high amplitude 
structures, which are 
all likely candidates when searching for units that exhibit extreme events. In addition, the island structures consist of multiple 
units, and a unit crossing the threshold often pulls one or both of its neighbours beyond the threshold. Combined, these two 
factors result in the extreme events happening in quick succession. Further, at low coupling strengths, it would seem that the 
oscillatory behaviour of the islands cannot appear in the statistics because the frequency of oscillation is much slower than the 
rate at which new events occur. However, moving into the higher coupling regime, one sees that there are fewer island structures, and 
so less candidates for extreme events. This allows for the oscillatory nature of the island structures to enter into the 
distributions. Of course, we still have the main peak due to the pulling effect mentioned above, but its amplitude is reduced because 
there are fewer island structures contributing to the short waiting times between consecutive events.

When considering the second definition, the obtained distributions are even more complicated (see Fig.~\ref{fig:stats_q}d). 
They exhibit a complex peak structure. Moreover, they seem to have heavier tails than the previously considered distributions. In fact, 
this waiting time distribution can be said to consist of two distinct regimes. The first regime is a correlated regime - the peak 
structure in Fig.~\ref{fig:stats_q}d, which takes place on short time scales, and is almost certainly related to the ability of chain 
units to temporarily retain 
energy allowing for a successive sequence of events. The second regime is an uncorrelated regime, taking place on long time scales (see 
Fig.~\ref{fig:log3}-3), where new events happen at seemingly random intervals. This part of the distribution has a decaying exponential 
dependence. The emergence of the peak structure is discussed in more detail in Sec.~\ref{sec:peaks}.

\subsection{Extreme Events: \texorpdfstring{$E$}{E}}
The distributions of maxima in energy $E$ for extreme events follow a similar behaviour to that of the amplitudes $q$ in that they 
exhibit an exponential dependence (see Fig.~\ref{fig:log3}-2). However, in 
contrast, there is a broadening of the distributions and an increase in the peak height with increasing $\kappa$. This suggests that 
the chain preserves (maybe even 
enhances) its ability to localise energy when the strength of the interactions increase. These changes are not as 
pronounced as in the case of the $q$ distributions. 

The distributions for event duration also follow similar behaviour to their $q$ counterpart, as the likelihood of long duration extreme 
events decreases with increasing $\kappa$. In contrast though, a peak structure emerges. This is related to the previously mentioned 
observation where the oscillations of the energy content of a unit which, while present, are often relatively small compared to 
oscillations in $q$. This effect enters the statistics because often the energetic oscillations happen above 
the threshold -- rather than crossing the threshold to end an extreme event, before recrossing it again to start another -- thus 
increasing the duration of events.

Looking at the time between events when using the first definition from the previous section (time 
between consecutive events occurring 
at any chain location), we see that the time decreases, with long waiting times becoming less likely for larger coupling strengths. 
This distribution is more regular than in the $q$ threshold case, as there appears to be just a single maximum peak located at 
zero.

However, these distributions become irregular when we consider 
the second definition for the waiting times between events (time between consecutive events occurring at each chain location), with 
additional peaks forming, and the tails becoming heavier. These complicated structures suggest that there are multiple 
mechanisms involved in the creation of high amplitude/energy 
fluctuations.

\section{Peak Structure: Discrete Breathers?}\label{sec:peaks}
\begin{figure}
  \vspace{0.1cm}\includegraphics[height=3.3cm, width=0.225\textwidth]{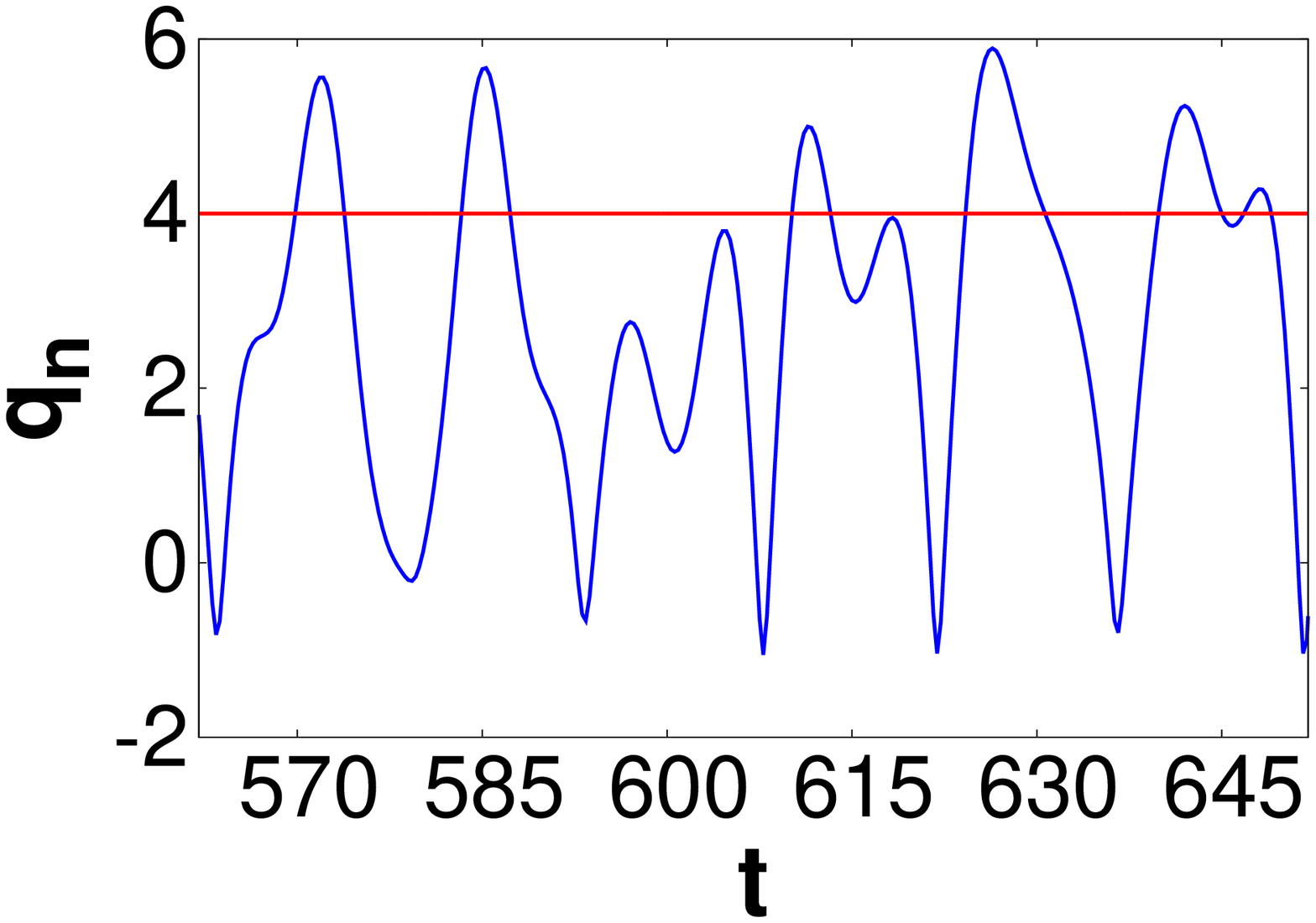}
  \includegraphics[height=3.3cm, width=0.25\textwidth]{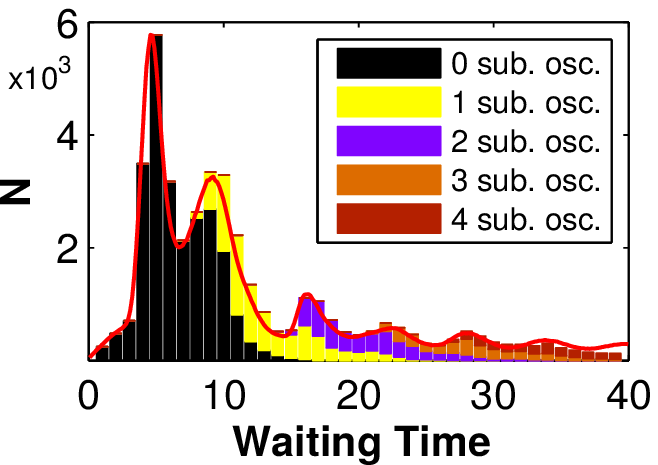}
  \caption{Color online. Left: Time series for a single chain unit, which crosses the threshold $q_n=4$ multiple times. Right: 
The stacked histogram shows the distribution of waiting times between consecutive extreme events when decomposed 
into the number of subthreshold oscillations between events. The red curve (solid line) shows the full distribution (from previous 
section) when no 
such decomposition has taken place. This curve has been normalised such that its largest peak coincides with 
that of the stacked histogram. In both panels $\kappa=0.2$.}\label{fig:crossings}
\end{figure}

\noindent
Examining a sequence of extreme events (in $q$) one sees that the localised structures oscillate in time. This oscillatory behaviour 
means that a single localised structure may cross the threshold $q_{thres}=4$ multiple times (see Fig.~\ref{fig:crossings} - left 
panel) resulting 
in multiple extreme events coming from a single localised structure.

It seems reasonable that this oscillatory behaviour may partially explain the interesting collection of peaks seen in the statistics of 
the previous section. Of course, the oscillatory behaviour is quite irregular. However, perhaps it is possible to predict the frequency 
of 
oscillation for these localised structures by exciting exact breathers of similar amplitude.

It is worth remembering that the local Morse potential is a soft anharmonic potential, which means that the larger the amplitude of the 
oscillation, the slower the frequency of oscillation. Therefore, the larger amplitude localised structures (that can be expected for 
low values of $\kappa$) should oscillate with a lower frequency than the localised structures with lower amplitude. The same is true 
for the breather solutions. To obtain such breathers numerically one is referred to \cite{marin2,flach2}.

Extracting the power spectrum from the system, for various values of the coupling strength $\kappa$, one sees that a wide range of 
frequencies are excited. These frequencies include the windows of frequencies that would correspond to discrete breathers. However, the 
corresponding powers do not suggest that these windows are in anyway more significant than any other part of the spectrum. Therefore, 
one must conclude that discrete breathers are not playing a key role in the system dynamics. A better way to characterise the 
evolution of these high amplitude structures would be to characterise them as multi-site chaotic (both spatially and temporally) 
breathers.

So if discrete breathers do not provide a solution to the emergence of the peaks that are present in the statistical distributions of 
Sec.~\ref{sec:stats}, then where do they come from? To find an answer to this question, it is worth considering the evolution of a 
single initial condition at a time when one of the units undergoes multiple successive extreme events (see Fig.~\ref{fig:crossings}). 
One thing that is apparent in this figure is that prior to each extreme event there is a number $k$ of subthreshold oscillations, i.e. 
oscillations whose amplitudes are less than $q_{thres}$, with $k \geq 0$. 
 
Taking a single initial condition and decomposing the waiting time statistics by the number of sub-threshold oscillations between 
consecutive extreme events one finds that the obtained distributions indeed exhibit a peak structure: Fig.~\ref{fig:crossings} - 
right panel. Moreover, this peak structure 
(from a  single initial condition) coincides extremely well with the peak structure in the distributions (from an ensemble of 
initial conditions) that are obtained when no such decomposition is carried out. 

\begin{figure}
  \includegraphics[height=3.3cm, width=8cm]{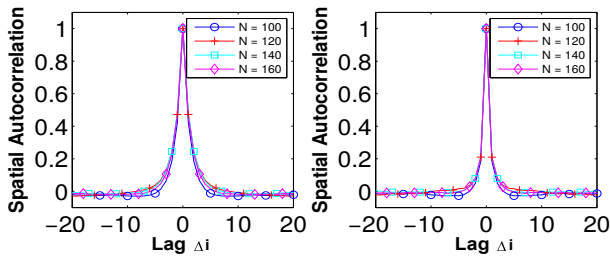}
  \caption{Color online. Spatial autocorrelation function for amplitudes $q$ (left) and energy $E$ (right), with different chain 
lengths, and with 
coupling strength $\kappa=0.1$.}\label{fig:acf}
\end{figure}

Before summarising our findings it is worth mentioning an aspect of our study that has gone largely unmentioned: the effects of 
chain length, relating to finite size effects, have not been 
considered, i.e how would chain length influence the statistical distributions? To answer this question, we computed the spatial 
autocorrelation function (see Fig.~\ref{fig:acf}). The sharp decay in correlations for all chain lengths lead us to believe that finite 
size effects are not an issue, and there would be no qualitative changes to any of the statistical distributions. 

\section{Conclusions \& Discussion}
We have studied a conservative Hamiltonian system modelling a one dimensional chain of harmonically coupled units. We have been 
particularly interested in the high amplitude and the high energy fluctuations that can result because of the coupled dynamics. Some of 
these 
fluctuations cross a threshold thus instigating what are termed extreme events. With respect to the dynamics, we observed that energy 
could move along the chain and become localised allowing for the formation of high energy island-like structures. Importantly, we have 
demonstrated that 
energy localisation should be considered in terms of a high energy background. These high energy island-like structures, possibly 
containing 
multiple units, can move coherently along the chain and influence the statistical properties of extreme events. 

A statistical analysis of the extreme events has unveiled some interesting properties of such events. For example, the distributions 
for amplitudes of extreme events exhibits a clear exponential dependence. In addition, the waiting times between consecutive events 
happening at the same chain location can be broken into regimes: the first a correlated regime where multiple events happen in quick 
succession due to a temporary retention of energy at some chain location, the second an uncorrelated regime where events happen 
seemingly at random. We have also demonstrated that the peaks in the distributions related to this first regime result because often 
there are a number of subthreshold oscillations separating two extreme events. 

Certainly it is of 
interest to extend the present study to one where different underlying potentials are considered. For example, using a hard or harmonic 
potential instead of the soft potential used here will result in dynamics that are both quantitatively and qualitatively 
different \cite{brown,reigada}. Thus, we can expect that the statistics of extreme events will also be different.

\section*{Acknowledgements}
SB acknowledges support by the Volkswagen Foundation (Grant No. 85390).

\bibliography{references}
\end{document}